\documentclass[aps,preprint,onecolumn,secnumarabic,nobalancelastpage,amsmath,amssymb,
nofootinbib]{revtex4}

\usepackage{enumerate}

\usepackage{graphicx}      
\usepackage{longtable}     
\usepackage{url}           
\usepackage{dcolumn}
\usepackage{bm}            
\usepackage{mathrsfs}
\usepackage{epstopdf}
\usepackage{hyperref}
\usepackage{color}
\usepackage[usenames,dvipsnames,svgnames,table]{xcolor}

\begin{document}


\title{Thermodynamics of rotating black holes in conformal gravity}

\author{Negin Kamvar${}^1$}
\author{Reza Saffari${}^1$}%
\email{rsk@guilan.ac.ir}
\author{Saheb Soroushfar${}^1$}
 \affiliation{${}^1$Department of Physics, University of Guilan, 41335-1914, Rasht, Iran.}%



\date{\today}

\begin{abstract}
In this paper we consider a metric of a rotating black hole in
conformal gravity. We calculate the thermodynamical
quantities for this rotating black hole including Hawking
temperature and entropy in four dimensional space-time, as we obtain
the effective value of Komar angular momentum. The result is valid
on the event horizon of the black hole, and at any radial distance
out of it. Also we verify that the first law of thermodynamics will
be held for this type of black hole.
\end{abstract}

\maketitle


\section{INTRODUCTION}
During the last century Einstein Gravity (EG) was one of the corner
stones of theoretical physics. Despite of the success in explanation
of various gravitational phenomena in nature, there are some
unsolved basic problems such as singularity problem, black hole
physics, and most importantly quantum theory of gravity. There
was an enormous effort in these lines to solve such problems but up
to now, it has not been obtained a complete theory of gravity. One
of such alternative theories of gravity is Conformal Gravity(CG)
\cite{Maldacena:1997re}. According to \cite{Fabbri:2011vk} Conformal Gravity which is an elegant theory because of its lagrangian constructed by Weyl tensor as a unique lagrangian. On the other hand, it is an important theory from phenomenological view point because of its scale symmetry relation to property of renormalizability, also it is intrested in some observational evidence such as, description of background effects of galactic rotation curve in the story of dark matter is discussed by Mannheim and Kazanas in \cite{P}. Intuitively, beside of local
Lorentz symmetry, it also has an scaling symmetry in which the
physics is invariant under changing the scale of the metric as
$g_{\mu\nu}\longrightarrow e^{\Omega(x)}g_{\mu\nu}.$ A
detailed introduction on conformal gravity is described
in \cite{Fabbri:2008ji}.

This paper is organized as follows: in section (\ref{Metric}), we introduce the
local solution of a neutral rotating black hole in pure conformal
gravity. The main purpose of this paper is calculating the
thermodynamical quantities for this type of black holes. Black hole
thermodynamics emerged from the classical general relativistic laws
of black hole mechanics, summarized by Bardeen - Carter - Hawking,
together with the physical insights by Bekenstein about black hole
entropy \cite{Jacob} and the semiclassical derivation by Hawking of
black hole evaporation. All the results that had been obtained from
1963 to 1973 culminated in the famous four laws of black hole
mechanics by Bardeen et al. \cite{Bardeen}; therefor, in section (\ref{thermodynamic}),
we obtain the thermodynamical quantities. We calculate some details
in evaluating the temperature, entropy and angular momentum, for the
case black hole of this paper. Hawking radiation results from the
quantum effect of fields in a classical geometry with an event
horizon. The flux of Hawking radiation can be also obtained through
the scattering analysis and there have been the studies of the grey
body factor for various black holes to calculate the Hawking
temperature $(T_{BH}=\frac{\kappa}{2\pi})$ \cite{ChangYoung:2010rz}.
To calculate the entropy of the black hole according to Bekenstiein
black hole entropy \cite{Jacob} there are a number of similarities
between black-hole physics and thermodynamics. Most striking is the
similarity in the behaviors of black hole area and of entropy. In
Wald formula for entropy had shown that this term is dependent on
area of the black hole $(S_{BH}=\frac{A}{4\pi G})$ \cite{Wald:1999vt}.
In this paper we show this fact with a different coefficient because
of the lagrangian density that we use in Wald integral. At the end
of this section we calculate the effective value of Komar angular
momentum by the space-like killing vector of black hole and using the
hodge operators. In section (\ref{first}), it is verified that the first law of
thermodynamics will be held. The paper is concluded in section (\ref{conclusion}).

\section{The metric}\label{Metric}
In this section, we briefly explain the metric solution of a black
hole in conformal gravity \cite{Sara}. Here we start with the action
\begin{align}\label{action}
S_{CG}=-\alpha_g\int d^{4}x\sqrt{-g}C_{\mu\nu\lambda\theta}C^{\mu\nu\lambda\theta}+S_M
=-2\alpha_g\int d^{4}x\sqrt{-g}[(R_{\mu\nu})^{2} - \frac{1}{3}R^{2}]+S_M,
\end{align}
where
$$C_{\mu\nu\lambda\theta}=R_{\mu\nu\lambda\theta}+\frac{1}{6}R
[g_{\mu\lambda}g_{\nu\theta} - g_{\mu\theta}g_{\lambda\nu}]
-\frac{1}{2}[g_{\mu\lambda}R_{\nu\theta} -
g_{\mu\theta}R_{\lambda\nu} - g_{\nu\lambda}R_{\mu\theta} +
g_{\mu\theta}R_{\nu\lambda}],$$ is the Weyl conformal tensor. As a
result the overall coupling of the theory $(-\alpha_g)$ is
dimensionless which seems is a good news for UV finiteness of the
theory \cite{Sara}. After varying the action with respect to the
metric and one obtains the equation of motion as
\begin{equation}\label{Bach tensor}
4\alpha_g W^{\mu\nu}=T^{\mu\nu}_{M},
\end{equation}
where $T^{\mu\nu}_{M}$ is Bach tensor and is defined as
$$ W^{\mu\nu}=\frac{1}{3}\nabla_{\mu}\nabla_{\nu}R
- \nabla_{\lambda}\nabla^{\lambda}R_{\mu\nu}
+\frac{1}{6}(R^{2}+\nabla_{\lambda}\nabla^{\lambda}R -
3(R_{\kappa\theta})^{2})g_{\mu\nu}
+2R^{\kappa\theta}R_{\mu\kappa\nu\theta} - \frac{2}{3}RR_{\mu\nu}.$$
In addition, one finds that the matter part of the action should
also respect to the scaling symmetry because the left hand side of
the above equation is traceless so the matter part of the action
should have a traceless energy-momentum tensor. Fortunately, by
introducing a conformal coupling term for the scalar mass term is
the standard model lagrangian is also conformably invariant
\cite{Mannheim:2011ds}. In particular it has obtained
\begin{equation}\label{Langrangian}
\mathcal{L}=\frac{1}{2}(D_{\mu}\phi)^{+}(D^{\mu}\phi)
- \frac{1}{12}R|\phi |^{2} - \frac{\lambda}{4}|\phi |^{4}
- \frac{1}{4}F^{a}_{\mu\nu}F^{a\mu\nu},
\end{equation}
where
$$ D_{\mu}=\nabla_{\mu} - ieA_{\mu}^{a}T_{a},$$
and $F^{a}_{\mu\nu}$ is the Lie algebra valued field strength tensor
of gauge field. After solving the equation of motion for these
fields it has also obtained \cite{Sara}
\begin{equation}
T^{\mu\nu}_{M}=\frac{1}{6}[g^{\mu\nu}\nabla^{\lambda}\nabla_{\lambda}|\phi |^{2}
- \nabla^{\mu}\nabla^{\nu}|\phi |^{2} - G^{\mu\nu}|\phi |^{2}],
\end{equation}
where $ G^{\mu\nu} $ is Einstein tensor.

 \subsection{\textbf{Rotating Black hole}}

In this part we use the slowly rotating solutions for pure conformal
gravity, that obtained in \cite{Sara}. Let us consider the following
line element around a rotating black hole
\begin{equation}\label{metric}
ds^{2}=\beta(r) dt^{2} - \frac{dr^{2}}{\beta(r)} - r^{2} d\theta^{2}
- r^{2}\sin^{2}\theta (d\phi - \frac{N(r)}{r} dt)^{2},
\end{equation}
where
$$\beta(r)=C_1+\frac{1}{3}\frac{C_1^{2}-1}{C_2r}+C_2r+C_3r^{2},$$
and
$$N(r)=C_4,$$
where $C_{1}=\sigma$ here we consider as constant of integration,
$\frac{C_1^{2}-1}{C_2}=-m$, $C_{2}$ is the coefficient that appears
in metric because of the Conformal Gravity solution, $C_{3}=-\frac{\lambda}{3}$,
that $\lambda$ is the cosmological constant, and $N(r)$ is the
constant value independent to r ,$(N(r)=\omega)$;therefor, we can
write the metric solution as a familiar form of

\begin{equation}\label{beta}
\beta(r)=\sigma-\frac{1}{3}\frac{m}{r}+Cr-\frac{\lambda}{3}r^{2}.
\end{equation}

\section{thermodynamical quantities}\label{thermodynamic}
In this section, we calculate the thermodynamical quantities of a
rotating black hole with the metric in previous section.  We work in
a system, that the value of $\hslash=G= C=1$.

\subsection{\textbf{Singularity and area of the event horizon}}

In this part, first we obtain the black hole singularity by solving
the equation $\beta(r)=0$, so we can find the radius of black hole.
This is a cubic equation that have three roots for ${r}$. Two of the
roots are imaginary and for this reason they will be neglected. The
other one is positive and the largest(${r_+}$) and it gives the
physical information that we want to obtain in this paper. After
that, we obtain the area of the horizon for the black hole, which is
a considerable importance because of the area theorem, which states
that  the horizon area of a classical black hole can never decrease
in any physical process.

By setting $dr=dt=0$ in the metric line elements, we can find line
elements for the 2-Dimensional horizon,

\begin{equation}\label{horizon metric}
d\sigma^{2}=-r_+^{2}d\theta^{2} - r_+^{2}\sin^{2}\theta d\phi .
\end{equation}

The area of the black hole horizon is then
\begin{equation}\label{horizon}
A=\int^{2\pi}_{0}d\phi \int^{\pi}_{0}\sqrt{|\det\gamma |} d\theta ,
\end{equation}

where $\gamma$ is the metric tensor for the black hole horizon.

\subsection{\textbf{Entropy}}

The entropy of black holes can be computed by the Wald formula
\cite{Wald:1999vt}

\begin{equation}\label{entropy}
S=-8\pi\int_{r=r_{+}}\sqrt{h}d^{2}x\epsilon_{ab}
\epsilon_{cd}\frac{\partial\mathcal{L}}{\partial R_{abcd}} ,
\end{equation}
where $h$ is the metric determinant on the surface.

In conformal gravity we have \cite{Liu:2012xn}

\begin{equation}\label{Conformal Langrangian}
\mathcal{L}=\frac{1}{2}\alpha C^{2}=\frac{1}{2}\alpha
(R^{\mu\nu\rho\sigma}R_{\mu\nu\rho\sigma}
- 2R^{\mu\nu}R_{\mu\nu}+\frac{1}{3}R^{2}) .
\end{equation}

After some calculations one can show that

\begin{equation}\label{CL}
\mathcal{L}=\frac{1}{2}\alpha(R_{\mu\nu}R^{\mu\nu} - \frac{1}{3}R^{2}) .
\end{equation}

The indices $a$ and $b$ take the ($t$,$r$) directions, thus we have

\begin{equation}\label{Langrangian derivative}
\epsilon_{ab}\epsilon_{cd}\frac{\partial \mathcal{L}}{\partial R_{abcd}}
=\frac{1}{2}\alpha[-(g^{rr}R^{tt}+g^{tt}R^{rr})+\frac{2}{3}g^{tt}g^{rr}R] ,
\end{equation}
and
\begin{equation}\label{horizon integral elements}
d^{2}x\sqrt{h}\vert_{r=r_{+}}=d\theta d\phi
\sqrt{g_{\theta\theta}g_{\phi\phi}} \vert_{r=r_{+}},
\end{equation}
therefore by using Eq.(\ref{Langrangian derivative}) and Eq.(\ref{horizon integral elements})we find the entropy as follow
\begin{equation}\label{entropy1}
S=4\pi \alpha[\frac{4m}{9r_{+}^{3}}-\frac{4}{3r_{+}^2}(1+\sigma)
-\frac{2C}{r_{+}}+\frac{2}{9}\lambda(1+3r_{+})+\frac{\omega^{2}}{3\pi r_{+}^{2}}]A_{H} ,
\end{equation}
where $A_{H}$ is the area of the black hole.

\subsection{\textbf{Temperature}}

In this part we attempt to obtain temperature of the aforementioned
black hole. According to the Hawking radiation theorem, black hole
temperature is dependent on surface gravity$(\kappa)$, that it is
equal to \cite{Oyvind}

\begin{equation}\label{surface gravity}
\kappa=\lim _{r\rightarrow r_{+}}\frac{\sqrt{a_{\mu}a^{\mu}}}{u^{t}} ,
\end{equation}
where
\begin{equation}\label{4-vector acceleration}
a^{\mu}=\Gamma^{\mu}_{\nu\lambda}u^{\nu}u^{\lambda}
=(u^{t})^{2}(\Gamma^{\mu}_{tt}+2\Omega_{H}\Gamma^{\mu}_{t\varphi}
+\Omega_{H}^{2}\Gamma^{\mu}_{\varphi\varphi}),
\end{equation}

in which $\Omega_{H}$, is the angular velocity of the black hole and
equal to

$$\Omega_{H}=-\frac{g_{t\varphi}}{g_{\varphi\varphi}}.$$

The normalization condition verifies that
\begin{equation}\label{normalization}
1=u^{\mu}u_{\mu}=(u^{t})^{2}(g_{tt}
+2\Omega_{H}g_{t\varphi}+\Omega^{2}g_{\varphi\varphi}).
\end{equation}

So we obtain $a^{\mu}a_{\mu}$ as
\begin{equation}\label{4-vector acceleration1}
a^{2}=a^{\mu}a_{\mu}=|g^{rr}|(\partial_{r}\ln u^{t})^{2}
+|g^{\theta \theta}|(\partial_{\theta}\ln u^{t})^{2},
\end{equation}
where $u^{t}=\frac{1}{\beta(r)}$. Using the inverse metric
coefficient
\begin{equation}\label{gr}
g^{rr}=-\beta (r),
\end{equation}
\begin{equation}\label{g}
g^{\theta \theta}=-\frac{1}{r^{2}}.
\end{equation}

Since we are interested to obtain the surface gravity on the event
horizon, so only we calculate the first term of Eq.(\ref{4-vector acceleration1}), then as a
result the surface gravity on the event horizon is equal to

\begin{equation}
\kappa=\frac{1}{2}\beta^{\prime}(r_{+}),
\end{equation}
where
\begin{equation}
\beta^{\prime}(r_{+})=C_{2} -\frac{1}{3}\frac{ C_{1}^{2} - 1}{C_{2}r_{+}^{2}} + 2C_{3}r_{+}.
\end{equation}

by using Eq.(\ref{4-vector acceleration1}) and Eq.(\ref{gr}) we can calculate the surface gravity
\begin{equation}
\kappa=\frac{1}{2}(\frac{1}{3}\frac{m}{r^{2}}-\frac{2}{3}\lambda r+C),
\end{equation}
thus the temperature is given by
\begin{equation}
T_{H}=\frac{\kappa}{2\pi}.
\end{equation}\\

\subsection{\textbf{Angular momentum}}

The Komar definition of the conserved quantity, corresponding to the
space-like Killing vector $\xi^{\mu}_{(\varphi)}$, in a coordinate
free notation is given by \cite{Modak:2010fn}

\begin{equation}\label{K}
K_{\eta}=\frac{1}{16\pi}\int *d\eta ,
\end{equation}
where
\begin{equation}\label{derivative}
d\eta=\frac{\partial g_{03}}{\partial r} dr\wedge dt
+\frac{\partial g_{03}}{\partial \theta} d\theta \wedge dt
+\frac{\partial g_{33}}{\partial r} dr\wedge d\varphi
+\frac{\partial g_{33}}{\partial\theta} d\theta \wedge d\varphi .
\end{equation}

Instead of working with $ dt,dr,d\theta, d\varphi$ we work with
orthonormal one forms, so we write Eq.(\ref{derivative}) as,

\begin{equation}
d\eta=\lambda _{10}\widehat{x_{1}}\wedge\widehat{x_{0}}
+\lambda_{20}\widehat{x_{2}}\wedge\widehat{x_{0}}
+\lambda_{13} \widehat{x_{1}} \wedge\widehat{x_{3}}
+\lambda_{23}\widehat{x_{2}}\wedge\widehat{x_{3}},
\end{equation}
where
\begin{equation}\label{lambda}
\nonumber \lambda _{10}=-\frac{\partial g_{03}}{\partial r}
-\frac{N(r)}{r}\frac{\partial g_{33}}{\partial r},
\end{equation}

\begin{equation}
\nonumber \lambda_{20}=-\frac{1}{r\sqrt{\beta(r)}}\frac{\partial g_{03}}{\partial \theta}
-\frac{N(r)}{r}\frac{\partial g_{33}}{\partial \theta},
\end{equation}

\begin{equation}
\nonumber \lambda_{13}=\frac{\sqrt{\beta(r)}}{r\sin(\theta)}\frac{\partial g_{33}}{\partial r},
\end{equation}

\begin{equation}
\lambda_{23}=\frac{1}{r^{2}\sin(\theta)}\frac{\partial g_{33}}{\partial \theta}.
\end{equation}

The dual of Eq.(\ref{derivative}) is \cite{Theodore}
\begin{equation}\label{dual}
*d\eta=\lambda_{10}\widehat{x_{2}}\wedge\widehat{x_{3}}
+\lambda_{20}\widehat{x_{1}}\wedge \widehat{x_{3}}
-\lambda_{13} \widehat{x_{2}}\wedge\widehat{x_{0}}
-\lambda_{23}\widehat{x_{1}}\wedge\widehat{x_{0}} .
\end{equation}

We can write Eq.(\ref{dual}) as
 \begin{equation}
*d\eta= \mathcal{\delta}_{rt}dr\wedge dt+\mathcal{\delta}_{\theta t}d\theta\wedge dt
+\mathcal{\delta}_{r \varphi}dr\wedge d\varphi + \mathcal{\delta}_{\theta \varphi} d\theta\wedge d\varphi ,
 \end{equation}
 where
 \begin{equation}
 \nonumber \delta_{\theta \varphi}=\lambda_{10}r^{2}\sin \theta,
 \end{equation}
 \begin{equation}
 \nonumber \delta_{\theta t}=-\lambda_{10}rN(r)\sin \theta,
 \end{equation}
 \begin{equation}
 \nonumber \delta_{r\varphi}=\lambda_{20}\frac{r\sin\theta}{\sqrt{\beta (r)}},
 \end{equation}
 \begin{equation}
 \nonumber \delta_{rt}=-\lambda_{20}\frac{N(r)\sin \theta}{\sqrt{\beta (r)}}+\lambda_{23},
 \end{equation}
 \begin{equation}
  \delta_{\theta t}=\lambda_{13}r\sqrt{\beta(r)}.
 \end{equation}

To calculate komar effective angular momentum we need to define a
boundary surface $(\partial \Sigma)$ , that it is characterised by a
constant $r$ and $dt=-\frac{g_{03}}{g_{00}}d\varphi$, so we have

\begin{equation}
*d\eta=-\frac{g_{03}}{g_{00}}\mathcal{\delta}_{\theta t}d\theta\wedge dt
+\mathcal{\delta}_{\theta \varphi}d\theta\wedge d\varphi,
\end{equation}
so we can write Eq.(\ref{K}) as,
\begin{equation}\label{K integral}
K_{\eta}=-\frac{1}{16\pi}\int \frac{g_{03}}{g_{00}}
\mathcal{\delta}_{\theta t}d\theta dt+\frac{1}{16\pi}
\int \mathcal{\delta}_{\theta \varphi}d\theta\ d\varphi,
\end{equation}
Moving along a closed contour, the first term of the right hand side
gives the shift of time between the initial and the final events.
Since we are performing an integration over simultaneous events this
term must be subtracted from Eq.(\ref{K integral}) \cite{Cohen},\cite{Moses}, so we
write that as follow
\begin{equation}
K_{\eta}=\frac{1}{16\pi}\int \lambda_{10}r^{2}\sin \theta d\theta\ d\varphi.
\end{equation}
By using Eq.(\ref{lambda})
\begin{equation}\label{K integral1}
K_{\eta}=\frac{1}{16\pi}\int (-\frac{\partial g_{03}}{\partial r}
-\frac{N(r)}{r}\frac{\partial g_{33}}{\partial r})r^{2}\sin \theta d\theta d\varphi ,
\end{equation}
using the metric coefficient
\begin{equation}\label{g_{03}}
g_{03}=rN(r) \sin^{2}\theta,
\end{equation}
\begin{equation}\label{g_{33}}
g_{33}=-r^{2}\sin^{2}\theta.
\end{equation}
After calculating the integral in Eq.(\ref{K integral1}) by using Eq.(\ref{g_{03}}), Eq.(\ref{g_{33}}) we obtain
the angular momentum as below

\begin{equation}
J=\frac{1}{6}r^{2}\omega .
\end{equation}

\section{First low of thermodynamics}\label{first}

For perturbations of stationary black holes, the change of energy is
related to change of area, angular momentum and electric charge
according to equations below
\begin{equation}\label{TdS}
TdS=dE-dW ,
\end{equation}
where
\begin{equation}\label{dW}
dW=\Omega_{BH}dJ+\Phi_{BH}dQ .
\end{equation}

Since entropy is dependent to the area of black hole, thus $dS$ is
proportional to $dA$ :in addition, due to the energy of black hole
is dependent on it's mass, $dE$ is proportional to $dM$; as a
result, for Eq.(\ref{TdS}) we have

\begin{equation}
dM=\frac{\kappa}{8\pi}dA+\Omega_{BH}dJ+\Phi_{BH}dQ .
\end{equation}

For this black hole we have $\Phi_{BH}=0$ because it is neutral, and
$\Omega_{BH}=\frac{\omega}{r}$. Therefore the first law of
thermodynamics for this black hole is as follow

\begin{equation}
dM=\frac{\beta^{\prime}(r_{+})}{16\pi}dA +\frac{\omega}{r}dJ .
\end{equation}

In conclusion we saw that the first law of black holes
thermodynamics is held.

\section{conclusion}\label{conclusion}

In this paper we have used the metric of a rotating black hole, that
has obtained in conformal gravity to calculate the thermodynamical
quantities of it. We have calculated the Hawking
temperature$(T_{BH}=\frac{\hslash\kappa}{2\pi})$ by the formula
$(\kappa=\lim _{r\rightarrow
r_{+}}\frac{\sqrt{a_{\mu}a^{\mu}}}{u^{t}})$ and the entropy of a
rotating black hole as the function of area of the black hole by
using the lagrangian density for the metric in conformal gravity
according to wald formula and after that we have calculated the
effective value of angular momentum with Komar expression for this
black hole at any distance $r$, by choosing boundary of a finite
spatial surface of radius $r$. This choice enabled us to evaluate
the Komar integrals without any asymptotic approximation. At the end
we have shown that the first law of the thermodynamics is held for
this black hole.

\bibliographystyle{amsplain}

\end{document}